\documentclass[12pt]{article}
\usepackage{amssymb,amsmath}

\newcommand{\la}{\lambda}
\newcommand{\om}{\omega}
\newcommand{\al}{\alpha}

\newcommand{\ga}{\gamma}

\newcommand{\fr}{\frac}

\newcommand{\cm}{\mu}

\newcommand{\cF}{\lambda}

\newcommand{\lb}{\label}

\newcommand{\be}{\begin{equation}}
\newcommand{\ee}{\end{equation}}
\newcommand{\ba}{\begin{eqnarray}}
\newcommand{\ea}{\end{eqnarray}}

\begin{document}
\begin{center}
{\Large\bf Form-field bremsstrahlung under collision of p-branes}\\

\bigskip
{\bf Dmitry Galtsov, Elena Melkumova and Karim Salehi}\\\medskip
       {\it Department of Physics, Moscow State University,
        Russia}
        \footnote{Talk given at  EPS International Europhysics Conference On High Energy Physics (HEPP-EPS 2005) ,
 21-27 Jul 2005, Lisbon, Portugal,\\
        E-mail: elenamelk@mtu-net.ru,
        E-mail: galtsov@phys.msu.ru}
        \end{center}
\begin{abstract} We calculate classically the radiation of the
antisymmetric form field generated in the collision of two
non-excited membranes moving with ultrarelativistic velocities in
five space-time dimensions. The interaction between branes through
the form field is treated perturbatively with the deflection angle
as a small parameter. Radiation arises in the second order
approximation if the domain of the minimal separation between the
branes moves with the superluminal velocity. It exhibits typical
Cherenkov cone features. Generalization to p-branes colliding  in
$D=p+3$ dimensions is straightforward. \end{abstract}
 \newpage
Recently  cosmological models involving strings and branes  moving
in higher-dimensional space-times received a renewed attention
\cite{DuKu05}-\cite{GaM01}. In particular, the problem of the
dimensionality of space-time can be approached within the brane
gas scenario \cite{DuKu05}-\cite{Ea01}. Another new suggestion is
the possibility of cosmic strings which are relics of the
super-strings in the early universe \cite{Po04}. In both cases it
seems necessary to reconsider the main mechanisms of radiation
which are applicable to extended objects like strings and branes.
In fact, it is well-known that dynamics of the usual cosmic
strings is essentially affected by radiation processes. The
radiation mechanism which has been mostly studied in the past
consists in formation of excited closed loops of strings which
loose their excitation energy emitting gravitons and axions.
Similar mechanism applies to other branes with axions being
replaced by appropriate antisymmetric forms.

Here we discuss another mechanism of radiation losses by colliding
branes which may work for the unexcited branes. It is similar to
bremsstrahlung of colliding charges in quantum electrodynamics. In
the simplest case of two parallel strings moving in
four-dimensional space-time the situation is equivalent to that of
colliding charges in 1+2 electrodynamics. More generally, strings
intersecting at an angle emit radiation if the velocity of the
point of their minimal separation (assuming strings to be moving
in two parallel planes) has a superluminal velocity (which if
infinite in the case of parallel strings). Radiation can be
interpreted as the Cerenkov effect \cite{GaGrLe93,GMK04}.

Similar mechanism works for colliding p-branes (of the same type)
in higher dimensions provided the brane codimension is equal to
two, i.e. $D-p=3$. The Cherenkov effect arises when the
hypersurface of the minimal separation of the dimension $p$ moves
superluminally. We concentrate on the case of membranes in
five-dimensional space time interacting via the rank three
antisymmetric form potential. Gravitational interaction is
neglected.

Consider two  classical membranes in the flat five-dimensional
(signature $-++++$) space-time $x^{\mu}=x_n^{\mu}(\sigma^a)$, $\mu
=0,1,2,3,4$, where $n=1,2$ is the brane label, interacting with
the three-form field $A_{\mu \nu \la}$. The world-volume of each
membrane is parameterized by the internal coordinates $\sigma^a
=(\tau , \sigma , \rho )$, and the signature is $+--$.

The total action consists of the membranes and the antisymmetric
form  terms $S=S_1+S_2+S_H$, where \be S_n= -
{\cm}_{n}\int\limits \sqrt{-\ga} \, d^3 \sigma \, - \fr{4\pi
{\cF}_n}{3!} \int\limits \epsilon^{abc}
\partial_a x ^ \mu \partial_b x ^ \nu \partial_c x ^ \la A_{ \mu \nu \la} \, d^3
\sigma,\ee  where $\gamma=\det \gamma_{ab}$ with
$\gamma_{ab}=\eta_{\mu\nu} {\partial_a x^\mu}
 {\partial_b x^\nu} $ being the induced
metric, $\epsilon^{012}=-1$, and \be S_H=
 \fr{1}{4!}\int\limits {H }_{ \mu \nu \rho \tau} { H}^{
\mu \nu \rho \tau } d^5 x, \quad { H}^{ \mu \nu \rho \tau
}=4!\partial_{[\mu}A_{\nu\rho\tau]}.\ee   Here ${\cm_n}$ and  $
{\cF}_n $ are the membranes tension parameters  and the coupling
constants. Using the world-volume diffeomorphism invariance, one
can impose three gauge conditions on the metric $\gamma_{ab}$, we
choose it to be diagonal, i.e. $
 \dot{x}^\mu x'_\mu  =  0, \;\; {\dot x}^\mu \bar{x}_\mu  =  0,
 \;\; x'^\mu\bar{x}_\mu  = 0$, where
$ \dot{x}^\mu  =  \partial_\tau x^\mu, \;
 x'^\mu = \partial_\sigma x^\mu, \ \bar{x}^\mu  =
  \partial_\rho x^\mu $.

The equations of motion for the membranes reads: \be \label{eq}
 \cm\; \partial_a  \left(\gamma^{ab}
\sqrt{-\gamma} \partial_b x^\mu  \right) = \fr{2 \pi {\cF}_n}{3}
V^{\alpha \beta \ga} \left(\partial^{\mu}A_{\al\beta
\ga}-3\partial_{\al}A^{\mu}_{\ \beta\ga}\right), \ee where the
world-volume form is $ V ^{\alpha \beta\ga} =  \epsilon^{abc}
\partial_a x^\al
\partial_b x^\beta \partial_c x^\ga $.
Here the three form is the sum of two contributions from membranes
$ A^{\mu\nu\la}=A^{\mu\nu\la}_1+A^{\mu\nu\la}_2 $, the field
equations in the Lorentz gauge $\partial_\mu A^{\mu\nu\la}=0$
being
 \be\label{eqA}
 \Box A_n^{  \mu\nu\la}=4\pi J_{n}^{ \mu\nu\la},\quad J_n^{\mu\nu\la}= \fr{{\cF}_n}{2}\int\limits V_n^{\mu
\nu\la} \delta^5 (x- x_{n}(\sigma))\,d^3 \sigma \ee
Our calculation follows the approach of~\cite{GMK04}. We solve the
branes equations of motion and the wave equation for the form
field iteratively in terms of the  coupling constant  ${\cF}$. In
the zero order approximation the branes are flat and moving
freely: $$\quad
   {}_0x^{\mu}_{n } \ = \ d^\mu_{n} \ + \ u^\mu_{n} \tau^n \
+ \Sigma^\mu_{n} \sigma^n + \ \Xi^\mu_{n} \rho^n ,$$ with the
impact parameter
 $d^\mu=d^\mu_2-d^\mu_1$, the 5-velocities $u^\mu_{n}$ and the
orientation vectors $\Sigma^\mu_{n}, \Xi^\mu_{n}$, with $\tau^n,
\,\sigma^n,\, \rho^n$ being the world-volume parameters on the
branes. The first brane is rest $ \quad u^ \mu_1 \ = \ (1,0,0,0,0)$
and assumed to lie in the 3-4 plane: $ \Sigma^ \mu_1 \ = \
(0,0,0,1,0), \quad \Xi^ \mu_1 \ = \ (0,0,0,0,1) $. The second brane
is inclined at the  angles $\al, \beta $ to the first one and moves
with the velocity $v$ orthogonal to the brane itself:
\begin{eqnarray}
   u^ \mu_2 &=& \gamma (1,0,-v \cos\al ,
v\sin \alpha \cos\beta, v \sin \alpha\sin\beta)  \nonumber\\
   \Sigma^ \mu_2 &=&
(0,0,\sin\al,\cos\alpha\cos\beta, \cos\alpha\sin \beta), \nonumber \\
   \quad\Xi^ \mu_2 &=& \ (0,0,0,-\sin\beta, \cos \beta),
\end{eqnarray}
  where $\gamma = (1-v^2)^{- \frac{1}{2}} $ and $ v \ = \
(1-(u_1u_2)^{-2})^{ \frac {1}{2}} $. If $d^\mu\neq 0$, the branes do
not intersect physically, while the line of their minimal separation
moves with the velocity $$V=\frac{v}{ \sin\alpha} .$$ Substituting
these parametrization
into (\ref{eqA}) one finds the first order form fields ${^{}_{1}A}%
_{n}^{\mu\nu\la} $  which describes mutual interaction between the
branes. This interaction gives rise to deformations of the branes
located mostly around the line of their minimal separation.
Deformations serve as a source of the second order fields
${^{}_{2}A}_{n}^{\mu\nu\la}$ which contain the radiative parts.
Radiation arises if the effective source moves superluminally,
i.e. $\sin \alpha<v$. It has typical Cherenkov angular
distribution determined by the equation $\omega= {\bf k}{\bf V}$,
which fixes two components  of the four-dimensional wave-vector
${\bf k}$ parallel to the brane at rest: $k^3 v=\omega \sin
\alpha\cos\beta,\;k^4 v=\omega \sin \alpha\sin\beta  $.

The radiation power can be computed as the rate of reaction
produced by the half sum of the retarded and advanced fields
upon the source and presented in the standard form \cite{GMK04}%
\begin{equation}
P\,^{0} = \ \frac1{6\pi^2}\int
k^{0}\epsilon(k^{0})|{_{1}J}_{\alpha\beta\ga}(k)|^{2}
\delta(k^{2})d^{5}k, \label{final}
\end{equation}
where $\epsilon$ is the sign function and the subscript 1
indicates that the current generating the radiation field results
from the first iteration order in the branes equations of motion.
A closed form expression for the  bremsstrahlung from the
collision of two branes can be obtained analytically in the case
of the
ultrarelativistic collision with the Lorentz factor $\gamma=(1-v^{2}%
)^{-1/2}\gg1$. The final formula for the spectral-angular
distribution of radiated energy per unit  area of the radiated
brane reads: \be \lb{Pkapbom}\fr1{ S}\fr{dP\,^0}{d\om
d\chi}=\fr{64\pi^4 {\cF}^6\kappa^2}{ {\cm}^2\om}
\fr{(1-\kappa^2\chi^2)^2}{(1+\kappa^2\chi^2)^4}
\exp\left({-\fr{\om d(1+\kappa^2\chi^2)}{\ga\kappa}}\right),\ee
where $\kappa=\gamma\cos\al$ and $ \chi= \fr\pi2 + \phi  $ and $
l_1 $ and $ l_2 $ are the length and $S$ is the normalization
area. The spectrum has an infrared divergence due to the
logarithmic dependence of the interaction potential between branes
with the  distance. Choosing as an infrared cutoff an inverse
length
 parameter $\Delta $ and integrating and over frequencies  we
 obtain the angular distribution of the total radiation
\be\lb{Pkapa} \fr1{ S}\fr{dP^0}{ d\chi}=\fr{64\pi^4 f^6\kappa^2}{
\mu^2} \fr{(1-\kappa^2\beta^2)^2}{(1+\kappa^2\beta^2)^4} {\rm  Ei}
\left(1, \fr{d(1+\beta^2\ga\kappa)}{\ga\kappa\Delta}\right) .\ee
Since the integral exponential function decays exponentially for
the values of the argument, the total radiation is peaked around
the direction $\chi=0$ $(\varphi=-\pi/2)$ within the angle
$\chi\lesssim \fr1{\sqrt{\ga\kappa}}$. This is conformal with the
Cerenkov nature of the effect.

Similar considerations apply to radiation of p-branes colliding at
a superluminal angle in $D=p+3$ dimensions (i.e. in the case of
branes of spatial codimension two). In this case the manifold of
the minimal separation has the spatial dimension $D-4$, and the
resulting expression for the radiated power differs only by the
normalization brane volume factor:
 \ba
\fr1{Vol_{D-3}}\fr{dP^0}{d\om d\chi}=\fr{64\pi^4
{\cF}_D^6\kappa^2}{ {\cm}_D^2\om}
\fr{(1-\kappa^2\chi^2)^2}{(1+\kappa^2\chi^2)^4}
\exp\left({-\fr{\om d(1+\kappa^2\chi^2)}{\ga\kappa}}\right).\ea

Other mechanisms of radiation from p-branes were considered in
\cite{AbCo00,GaM01,HaSa00}. The work was supported in part by the
RFBR grant 02-04-16949.

\end{document}